\documentclass[journal, two side]{IEEEtran}

\IEEEoverridecommandlockouts                              
\usepackage{epsfig} 
\usepackage{graphics} 
\usepackage{amsmath} 
\usepackage{amssymb}  
\usepackage{subfigure}
\usepackage{subeqnarray}
\usepackage{cite}
\usepackage{url} 
\usepackage{epstopdf}  

\begin{document}

\title{Battery Placement on Performance of VAR Controls}

\author{\IEEEauthorblockN{Hen-Geul Yeh and Son H. Doan}
\IEEEauthorblockA{Department of Electrical Engineering\\
California State University, Long Beach\\
Long Beach, CA 90840\\
Email: heyeh@csulb.edu, sonproee@gmail.com}
}

\markboth{Proceedings of Green Energy and Systems Conference 2013, November 25, Long Beach, CA, USA. }{This full text paper was peer reviewed at the direction of Green Energy and Systems Conference subject matter experts.}

\maketitle

\begin{abstract}
Battery's role in the development of smart grid is gaining greater attention as an energy storage device that can be integrated with a Photovoltaic (PV) cell in the distribution circuit. As more PVs are connected to the system, real power injection to the distribution can cause fluctuation in the voltage. Due to the rapid fluctuation of the voltage, a more advanced volt-ampere reactive (VAR) power control scheme on a fast time scale is used to minimize the voltage deviation on the distribution. Employing both global and local dynamic VAR control schemes in our previous work, we show the effects of battery placement on the performance of VAR controls in the example of a single branch radial distribution circuit. Simulations verify that having battery placement at the rear in the distribution circuit can provide smaller voltage variations and higher energy savings than front battery placement when used with dynamic VAR control algorithms.
\end{abstract}

\section{Introduction}
\label{sec:introduction}

Batteries in the electrical grid aren't just used as energy storage device, but also for providing many useful functions such as improving overall system efficiency and reducing greenhouse gas emissions \cite{Jewell2012}. They can be integrated with the renewable energy generators such as Photovoltaic (PV) and wind generators \cite{Zou2012}. When integrated with wind generators, they can help reduce the intermittency issue associated with wind availability from day to day \cite{Such2012,Grillo2012}. The integration of battery and renewable energy generators into the grid can take place both at the transmission or distribution level. Such integrations throughout a distribution circuit can pose a serious problem of injecting real power to the system that causes higher fluctuation in the voltage when the PV penetration level gets high enough. Advanced volt-ampere reactive (VAR) power control can be used to provide reactive power support, reduce power loss, and control the voltage levels through the distribution lines.

Capacitor banks placed in optimal location have been historically known to maintain  desired level of VAR draw from the substation circuit. However, both capacitors and under load tap changers (ULTCs) are not fast enough to compensate for transient events like clouds passing over the cells \cite{Farivar2011}. On the other hand, DC/AC inverters that connect PV or batteries to the grid however can serve as additional controllers because they can pull or push reactive power at a much faster timescale and with a much finer resolution than current technologies such as shunt capacitors and ULTCs \cite{Farivar2011}. Farivar et al. \cite{Farivar2011} propose a hybrid scheme without integrated energy storage for controlling both the inverters and shunt capacitors by imposing two different time scales. Turitsyn et al. have proposed a number of schemes focused on inverter power control without integrated energy storage \cite{Turitsyn2010a,Turitsyn2010b}. They first studied the problem of minimizing power losses \cite{Turitsyn2010a} and then examined a multi-objective optimization problem aimed at minimizing both power losses and the maximum voltage change along the circuit \cite{Turitsyn2010b}. DC/AC inverters are not the only factor that's currently being utilized for voltage regulation. Batteries, when integrated with renewable energy generators, have been shown to also mitigate the voltage variation on the grid. The work in \cite{Shao2013} studied the effect of distributed energy storage in distribution involving Plug-in Electric Vehicle (PEV). They showed that the use of energy storage could smooth demand fluctuations and mitigate feeder voltage fluctuation \cite{Shao2013}. Barnes et al. \cite{Barnes2012} also focused their work on the usage of energy storage coordinated with PV inverters for peak shaving and voltage regulation on a distribution system having a high level of PV penetration. In a distribution with very high level of PV penetration, the choice of battery and PV placement is much limited. Moreover, high level of PV penetration is currently not very realistic due to the relatively high cost for installing PV system with energy storage device. Hence, we need to also consider cases of low or medium PV penetration levels where placement of battery and PV system may make a difference in terms of energy savings and voltage variations. We then extend our work in \cite{Yeh2012}, in which we  presented two dynamic control algorithms with integrated battery for reactive power control in a single line radial distribution system with two different battery placements: 1) the front battery placement where all PV-enabled nodes with battery are placed at the front of the distribution, and 2) the rear battery placement where all PV-enabled nodes with battery are placed at the rear of the distribution. Note that the substation is located at the beginning of the distribution line and is considered as the front end of the distribution line. The purpose of this paper is to compare the results of two different battery placements and examine their effect on VAR control for low, medium, and high levels of PV penetration. Our approach is to solve the optimization aimed at minimizing the total power losses over a day by using both local control and global VAR control algorithms. 

This paper is organized as follows. The next section describes the problem setting and provides the relevant power flow equations for a radial distribution circuit in our previous work \cite{Yeh2012}. Following that, both global and local control algorithms are presented. Then, some example test cases are described and the corresponding simulation results are discussed. Finally, conclusion and future work are presented.

\section{Problem Setup}
\label{sec:problemSetup}

\begin{figure}[thpb]
\centering
\includegraphics[width=0.5\textwidth,clip]{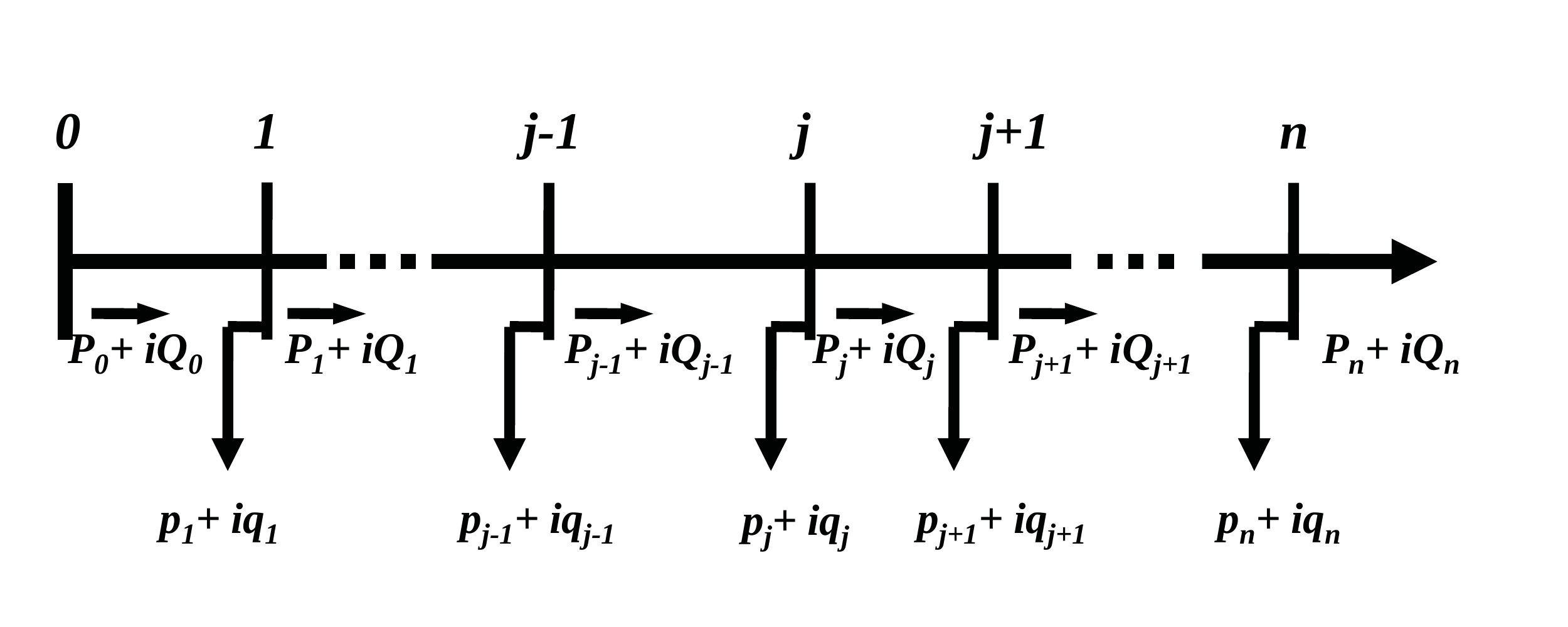}
\caption{Schematic diagram of a unidirectional single branch radial distribution circuit with $n+1$ nodes. For simplicity, the time variable $t$ is not shown.}
\label{fig:radial_diagram}
\end{figure}

As in \cite{Yeh2012} and \cite{Yeh2012b} , a unidirectional single branch radial distribution circuit is depicted in Figure \ref{fig:radial_diagram}. Node $0$ is located at the front-end (i.e. substation) and node $n$ is located at the rear-end. For simplicity, the discrete time variable $t$ is not shown in Figure \ref{fig:radial_diagram}, but shown in the remaining of the paper. Note that the time $t$ is a discrete time index in our model. It can be based on hours for one day. For example, if the time increment is one hour, then $t$ starts at 1 and ends at $T=24$. Or, it can also be based on faster time scales such as minutes for one day. For example, if the time increment $\Delta{t}$ is twenty-minute, i.e., $t$ starts at 1 and ends at $T=72$ (at the end of 24 hour).

For a unidirectional single branch radial distribution circuit, the power flows along the line from left to right. The symbols $P_{j}(t)$ and $Q_{j}(t)$ respectively represent the real and reactive power at each bus $j$ and at time $t$, where as the lowercase letters represents the net local node's real and reactive powers $p_{j}(t) = p^{c}_{j}(t) - p^{g}_{j}(t)$ and  $q_{j}(t) = q^{c}_{j}(t) - q^{g}_{j}(t)$, where $p^{c}_{j}(t)$ and $q^{c}_{j}(t)$ respectively denote the real and reactive power consumption at node $j$ and at time $t$; $p^{g}_{j}(t)$ and $q^{g}_{j}(t)$ represent the real and reactive power generated by the PV cells at node $j$ and at time $t$. The complex impedance in the line between nodes $j$ and $j+1$ (not shown) is denoted $r_j+\mathbf{i}x_j$.

The complex power flows at each node $j$ are described through the DistFlow equations from \cite{Baran1989c} for $j$ = $0,...,n-1$

\begin{equation}
\label{eqn:DistFlow01}
{P_{j + 1}(t)}={P_{j}(t)} - {r_j}\frac{{P_j^{2}(t) + Q_j^{2}(t)}}{{V_j^{2}(t)}} - p_{j + 1}^{c}(t) + p_{j + 1}^{g}(t) - \beta_j(t)
\end{equation}

\begin{equation}
\label{eqn:DistFlow02}
{Q_{j + 1}(t)} = {Q_{j}(t)} -x_j\frac{{P_j^{2}(t) + Q_j^{2}(t)}}{{V_j^{2}(t)}} - q_{j + 1}^{c}(t) + q_{j + 1}^{g}(t) - \eta_j(t)
\end{equation}


\begin{IEEEeqnarray}{rCl}
{V^2_{j + 1}(t)} & = & {V_{j}^{2}(t)} -2(r_j P_{j}(t)+x_j Q_{j}(t)) \nonumber\\
&& +(r_j^2+x_j^2)\frac{{P_j^{2}(t) + Q_j^{2}(t)}}{{V_j^{2}(t)}}
\label{eqn:DistFlow03}
\end{IEEEeqnarray}

\noindent
where $V_j(t)$ is the voltage at node $j$, at time $t$, $P_j(t)+\mathbf{i}Q_j(t)$ is the complex power flowing from node $j$ to node $j+1$ (i.e. from left to right in Figure \ref{fig:radial_diagram}), $p_j(t)+\mathbf{i}q_j(t) = (p_j^{c}(t) - p_j^{g}(t))  +\mathbf{i}(q_j^{c}(t)-q_j^{g}(t))$ is the complex power consumed at node $j$ and at time $t$, $\beta_j(t)$ denotes the rate of charge/discharge of real energy at time $t$, and $\eta_j(t)$ denotes the rate of charge/discharge of reactive energy at time $t$. For simplicity, $\eta_j(t)=0$. In general, the charging/discharging rate is restrained by the battery temperature rise and generated gas. However, for simplicity,  the charging/discharging rate (i.e., $\beta_j(t)$) is unrestricted and acts as one of the control variables in this paper. Each node's real and reactive power consumption $(p^c_j(t),q^c_j(t))$ and real power generation $p^g_j(t)$ are generally assumed to be known. Another control variable is the reactive power generation $q^g_j(t)$ (i.e., the reactive power supplied by the PV inverter). The control $q^g_j(t)$ can only be applied at the nodes with PV generation because there is no net power injection at the other nodes, i.e., at all non-PV nodes $p^g_j(t)$ and $q^g_j(t)$ are zero. The amount of real energy storage (battery) at node $j$ and time $t$ follows the simple difference equation 

\begin{equation}
\label{eqn:batteryEq}
b_j(t+1) = b_j(t) + \beta_j(t) \Delta{t},    
\end{equation}

\noindent
for $t$ = $1,...,T-1$ with initial condition $b_j(1)=0$.

At each bus $j$, the amount $b_j(t)$ of energy storage and the rate $\beta_j(t)$ of charging/discharging the battery per $\Delta{t}$ (unit timescale of 20 min) are computed in \eqref{eqn:batteryEq} and $b_j(t)$ is bounded as

\begin{equation}
\label{eqn:batteryBounds}
0 \le b_j(t) \le  B_j^{max},  
\end{equation}

\noindent
for $t$ = $1,...,T$, where $B_j^{max}$ is the maximum size of the battery that can be used as energy storage device in conjunction with DC/AC inverters. 

As previously discussed, control of the reactive power input at each node $q^g_j(t)$ can be efficiently accomplished using the local inverter of the PV generator. This work adopts the inverter model described in \cite{Turitsyn2010a} and \cite{Liu2008}. The inverter's ability to generate reactive power is constrained by its apparent power capability defined as $s_j(t) = |p^g_j(t)+\mathbf{i}q^g_j(t)|$ and its instantaneous real power generation $p^g_j(t)$ through the relationship

\begin{equation}
\label{eqn:reactivePowerEq}
\lvert q^g_j(t) \rvert \le \sqrt{s^2_j(t) - (p^g_j(t))^2} \equiv (q^g_j(t))^{max}  
\end{equation}

Given such a control input, the assumptions stated above, and either initial conditions (i.e., $j=0$) or terminal constraints (i.e., $j=n$) for the powers and the voltage; the DistFlow equations \eqref{eqn:DistFlow01}-\eqref{eqn:DistFlow03}  allow the computation of the power and voltage ($P_j(t),Q_j(t)$, and $V_j(t)$) at each node along the circuit. For our analysis, we have  $V_{0}(t)=1$ at  $j=0$ and $P_{n}(t)=Q_{n}(t)=0$ at $j=n$.
 
A natural control goal for distribution systems is to minimize or keep the power (rate of energy) loss along the circuit within an acceptable range. The total losses over a day along the entire distribution branch are given by

\begin{equation}
\label{eqn:opEq}
\mathcal{L} =  \mathop{\min} \sum\limits_{t = 1}^{T} \sum\limits_{j = 0}^{n-1} {r_j}\frac{{P_j^{2}(t) + Q_j^{2}(t)}}{{V_j^{2}(t)}}.
\end{equation}

We will assume that the state of the distribution network (the voltage, real and reactive power at each bus), and the input to the network (real and reactive power generated or consumed at each bus) remain unchanged within each time increment and may change only from $t$ to $t+1$ in a discrete manner when measured data are updated via sensors. The inverter control will be applied for each $t$ to match PV output fluctuations and the batteries will be served as a buffer in between PV cells and inverter; and the voltage will be adjusted for each discrete time to match load fluctuations. Another important consideration for proper operation of a distribution circuit is that the per unit voltage variation along the line remain within the bounds

\begin{equation}
\label{eqn:voltage_bounds}
1- \varepsilon \leq V_j(t) \leq 1+\varepsilon,
\end{equation}

Prior to introducing the final problem formulation, we discuss two simplifying assumptions. First, linearized power flow equations have been extensively used in the literature \cite{Turitsyn2010a,Turitsyn2010b,Baran1989c,Gilbert1998}. Their use was justified by Baran and Wu \cite{Baran1989c} based on the fact that the nonlinear terms represent the losses which in practice should be much smaller than the branch power terms $P_j(t)$ and $Q_j(t)$. Turitsyn et al. \cite{Turitsyn2010a} verified this assumption by showing that the nonlinear terms in \eqref{eqn:DistFlow01} are much smaller than the linear terms as well as a insignificant changes in the results for the linear versus nonlinear equations. Second, we also assume that $V_j(t) \approx V_0(t)$. The voltage approximation will be valid as long as the constraint \eqref{eqn:voltage_bounds} is met with a sufficiently small $\varepsilon$. Combining these two simplifications gives

\begin{equation}
\label{eqn:simpDistFlow01}
{P_{j + 1}(t)}={P_{j}(t)} - p_{j + 1}^{c}(t) + p_{j + 1}^{g}(t) - \beta_j(t)
\end{equation}

\begin{equation}
\label{eqn:simpDistFlow02}
{Q_{j + 1}(t)} = {Q_{j}(t)} - q_{j + 1}^{c}(t) + q_{j + 1}^{g}(t) 
\end{equation}

\begin{equation}
\label{eqn:simpDistFlow03}
{V_{j + 1}(t)} = {V_{j}(t)} - \frac{r_j{P_j(t) + x_jQ_j(t)}}{{V_0}}
\end{equation}

\noindent
along with a simplified expression for the total losses

\begin{equation}
\label{eqn:simpOpEq}
\mathcal{L} =  \mathop{\min} \sum\limits_{t = 1}^{T} \sum\limits_{j = 0}^{n-1} {r_j}\frac{{P_j^{2}(t) + Q_j^{2}(t)}}{{V_0^{2}}}.
\end{equation}

We refer  \eqref{eqn:simpDistFlow01}-\eqref{eqn:simpOpEq} as the simplified DistFlow branch equations. Given this expression, the system equations \eqref{eqn:simpDistFlow01}-\eqref{eqn:simpOpEq} and constraints described above, the problem of minimizing losses can be cast as an optimization problem

\begin{equation}
\label{eqn:optimization}
\mathop{\min}\limits_{P,Q,V,\beta,q^g} \mathcal{L}
\end{equation}

\noindent
subject to \eqref{eqn:batteryEq}, \eqref{eqn:batteryBounds}, \eqref{eqn:reactivePowerEq}, \eqref{eqn:voltage_bounds}, and \eqref{eqn:simpDistFlow01}-\eqref{eqn:simpDistFlow03}. We refer to this as the loss minimization problem. Loss minimization is often the only problem considered, however, reducing voltage variation along the circuit is also desirable.

\section{Control Algorithms}
\label{sec:cntrl}

The local control scheme is given as follows: the controller uses the optimal solution of the total loss minimization problem \eqref{eqn:simpOpEq} with the following local control constraint

\begin{equation}
\label{eqn:localControl}
q_j^g(t) := \left\{ \begin{array}{lr}
q_j^c(t),&{\rm{if}}\left| {q_j^c(t)} \right| \le (q_j^g(t))^{max} \\
{{\rm sgn} (q_j^c(t))(q_j^g(t))^{max}} ,&{\rm{ otherwise}},
\end{array} \right.\end{equation}

\noindent
which is based on the controller described in \cite{Turitsyn2010b} and leads minimization of the objective function \eqref{eqn:optimization} subject to  \eqref{eqn:batteryEq}, \eqref{eqn:batteryBounds}, \eqref{eqn:voltage_bounds}, \eqref{eqn:simpDistFlow01}-\eqref{eqn:simpDistFlow03}, and \eqref{eqn:localControl}. Note that the local control law employs two parameters $q_j^c(t),(q_j^g(t))^{max}$ or one measurement $q_j^c(t)$ to determine $q_j^g(t)$. In general, more measurement means high communication requirements.

The global control scheme uses the optimal solution of the total loss minimization problem \eqref{eqn:simpOpEq} with the control law that determines $q_j^g(t)$ from constraint \eqref{eqn:reactivePowerEq}, which is based on the inverter model described in \cite{Turitsyn2010a} and \cite{Liu2008}, and leads minimization of the objective function \eqref{eqn:optimization} subject to \eqref{eqn:batteryEq}, \eqref{eqn:batteryBounds}, \eqref{eqn:reactivePowerEq}, \eqref{eqn:voltage_bounds}, \eqref{eqn:simpDistFlow01}-\eqref{eqn:simpDistFlow03}.  The results of these two control laws will be discussed in greater details in Section \ref{sec:resultDiscussion}.

\section{Test Circuit Simulations}
\label{sec:testSetup}

\begin{figure}[thpb]
      \centering
      \includegraphics[width=0.5\textwidth,clip]{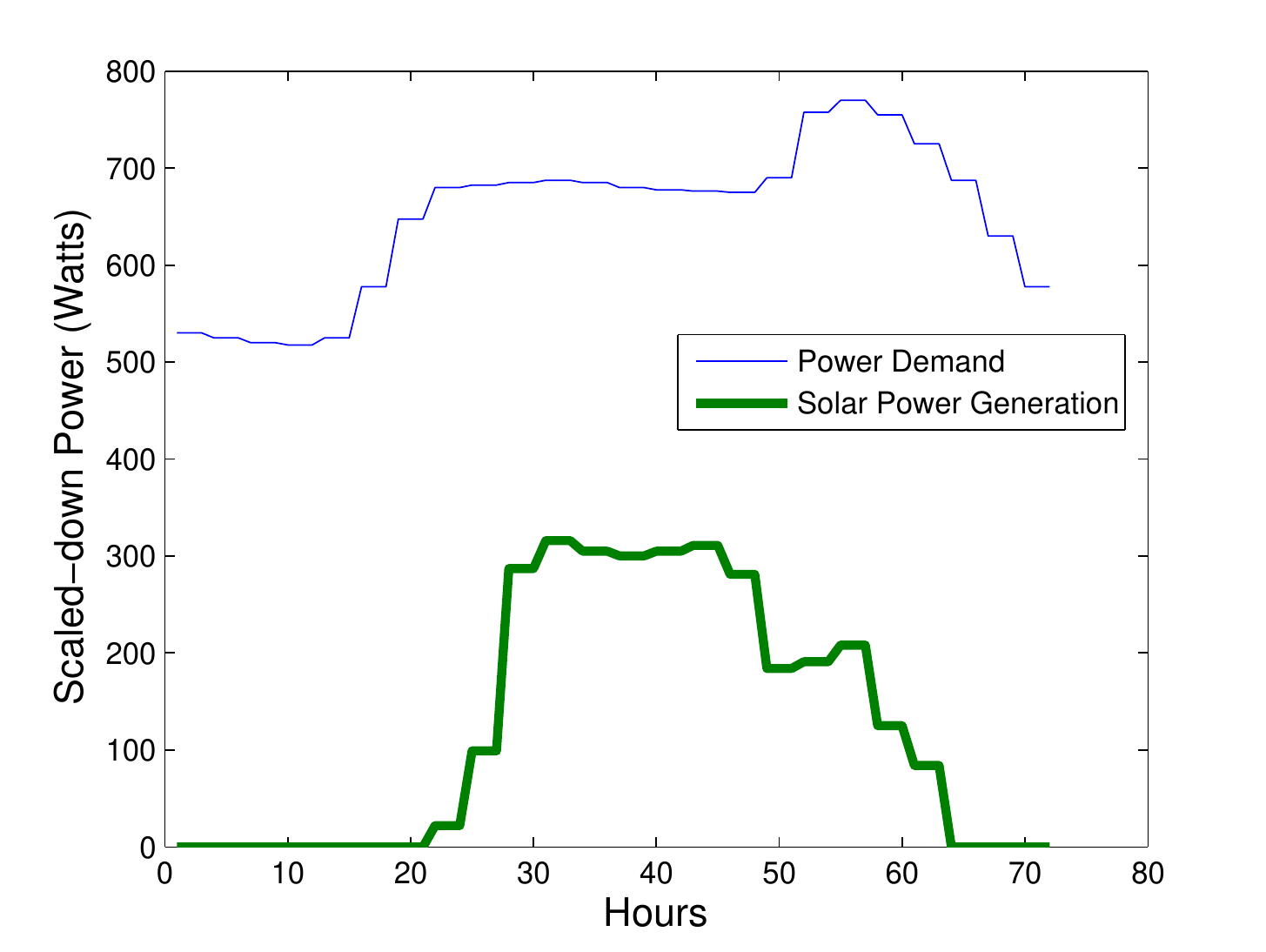}
      \caption{Power demand and generation profile from CAISO on Oct. 27,2011.}
      \label{fig:powerProfile}
\end{figure}

We focus on radial distribution networks and perform a deterministic approach to simplify the simulations. Although the circuit model is unidirectional, we don't place any constraints on the direction of the power and allow it to be both positive or negative. Therefore, this model can be used in the case of bidirectional circuit. The test circuits that we study in this paper are representative of rural distribution circuits. In particular, those with a three phase 12.47 kV nominal voltage, which corresponds to a single phase equivalent line to-neutral voltage of 7.2 kV at the sub-station. The simulations considered herein assume a circuit with 30 nodes separated by the distances drawn from a uniform distribution between 200 to 300 meters. The line impedances are all set to $0.33 + \mathbf{i}0.38$ Ohms/km, which is based on typical conductor types and sizes discussed in \cite{Schneider2008} and used in \cite{Turitsyn2010a}. Both scaled down power demand and generation profiles from CAISO on Oct 27, 2011, are employed as $q_j^c(t)$ and $p_j^g(t)$ and shown in Figure \ref{fig:powerProfile}. The scaled down factor applied for power demand is 40e6 and for solar generation is 1e6. Since CAISO only provide the data for each hour and 20-minute is the selected time increment, we set the same value for three time slots in each hour as depicted in Figure \ref{fig:powerProfile} for 24 hours ($x$-axis: 1 to 72 time slots).

We set $p_j^g(t)=(p_j^g(t))^{max}$ at all the PV-enabled nodes and the apparent power capability of the inverters $s_j(t)$ to a constant value $s_j(t) = s_{max}(p_j^g(t))^{max}$ (i.e. $s_{max}=1.1$ implies that the apparent power capacity of the inverters is $10\%$ greater than the real power capacity of the PV cells).  From \cite{avgHouseEnergy}, the monthly electricity consumption for a U.S. residential utility customer was $940$ kWh, which is equivalent to daily electricity consumption of about $31.33$ kWh. Assuming $B^{max}$ is $\frac{1}{20}$ of the daily residential consumption and using the same scaled down factor that was used for power demand, we then have the average $B_j^{max} = B^{max} = 20 \times 10^{-4}$ Joules, which is a constant for all batteries at PV-enabled nodes.

For the non PV-enabled nodes, $p_j^g(t),q_j^g(t)$, and $s_j(t)$ are all zeros. In this simulation, we use PV penetration level of $20\%$, $50\%$, and $80\%$, which is respectively denoted as $a = 0.2$, $a = 0.5$, and $a = 0.8$ on the plots shown in Section \ref{sec:resultDiscussion}. All PV-enabled nodes are assumed to have energy storage (battery) and are placed one after another at either the front or rear of the distribution. Finally, we specify the voltage bounds in \eqref{eqn:voltage_bounds} based on $\varepsilon = 0.05$.

\section{Results and Discussion}
\label{sec:resultDiscussion}

The comparison of the results is based on the following three different control laws: 1) the global solution of the total loss minimization problem (i.e. \eqref{eqn:optimization} subject to \eqref{eqn:batteryEq}, \eqref{eqn:batteryBounds}, \eqref{eqn:reactivePowerEq}, \eqref{eqn:voltage_bounds}, \eqref{eqn:simpDistFlow01}-\eqref{eqn:simpDistFlow03}), 2) the optimal solution when the control law is constrained by the local rule \eqref{eqn:localControl}, and 3) the optimal solution without inverter control (i.e., $q_j^g(t)=0$ for all $j$). In the following discussion, we respectively refer to these cases as the “Global Solution”, the “Local Control Law” and “No Control”. In the local control scheme described in Section \ref{sec:cntrl} the determination of the control $q_j^g(t)$ only depends on the local information, whereas the objective function is global. The performance of the algorithms is evaluated based on energy savings and voltage variation. The comparisons are made based on the results of penetration levels $a=0.2$, $a=0.5$, and $a=0.8$ for each of the control algorithm in two battery placements: 1) the front battery placement where all PV-enabled nodes are placed at the front of the distribution, and 2) the rear battery placement where all PV-enabled nodes are placed at the rear of the distribution. The voltage variation is defined as follows:

\begin{equation}
\label{eqn:voltageVariation}
\delta {V} = {\rm{ }}\mathop {\max }\limits_j \left| {\frac{{{V_j(t)} - {V_0}}}{{{V_0}}}} \right|,
\end{equation}

\noindent
for all $t$.

The energy savings is computed as a percentage of the losses versus the baseline case, where the baseline case is when there is no control (i.e., $q_j^g(t)=0$ for all $j$) as follows:

\begin{equation}
\label{eqn:energySaving}
Saving = \frac{\mathcal{L}(Baseline) - \mathcal{L}(ControlMethod)}{\mathcal{L}(Baseline)}
\end{equation}

\noindent
where $\mathcal{L}(baseline)$ and $\mathcal{L}(ControlMethod)$ are the total loss of the baseline (i.e. No Control) over one day and the total loss of either the global or local control method over one day, respectively. When there is no control (i.e. baseline), the savings becomes “zero”.

In the following discussion, we study these performance metrics over $s_{max}$ with front and rear battery placement in the optimization problems which are solved using CVX, a package for specifying and solving convex programs \cite{Grant2008,cvx} in MATLAB. The simulation is done with a PV penetration of $a = 0.2$, $a = 0.5$, and $a = 0.8$. 

\begin{figure*}
\begin{center}
\subfigure[]{\label{fig:dV_smax_Front}
\includegraphics[height=2.45 in, width=3.3 in,clip]{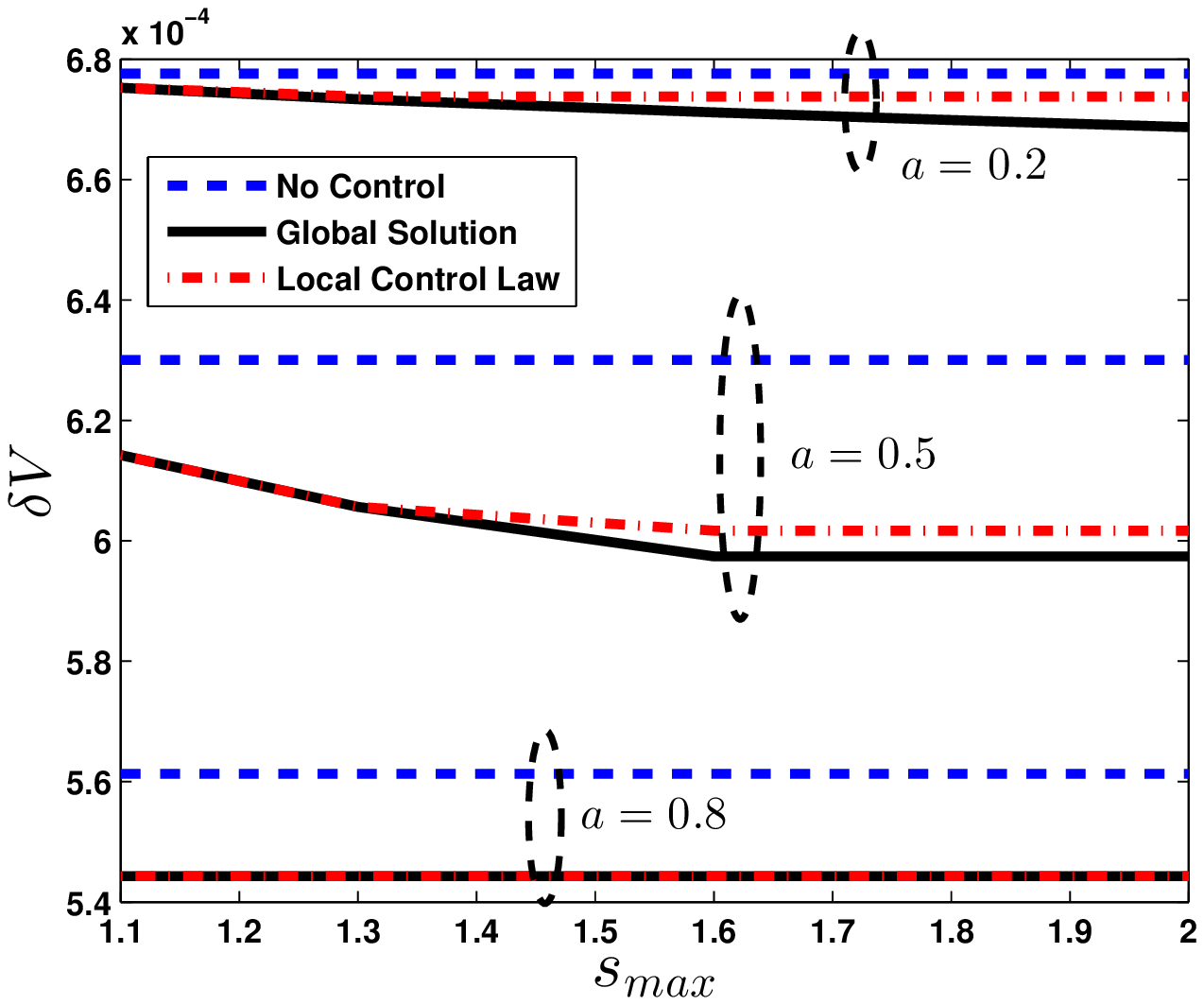}}
\subfigure[]{\label{fig:dV_smax_Rear}
\includegraphics[height=2.45 in, width=3.3 in,clip]{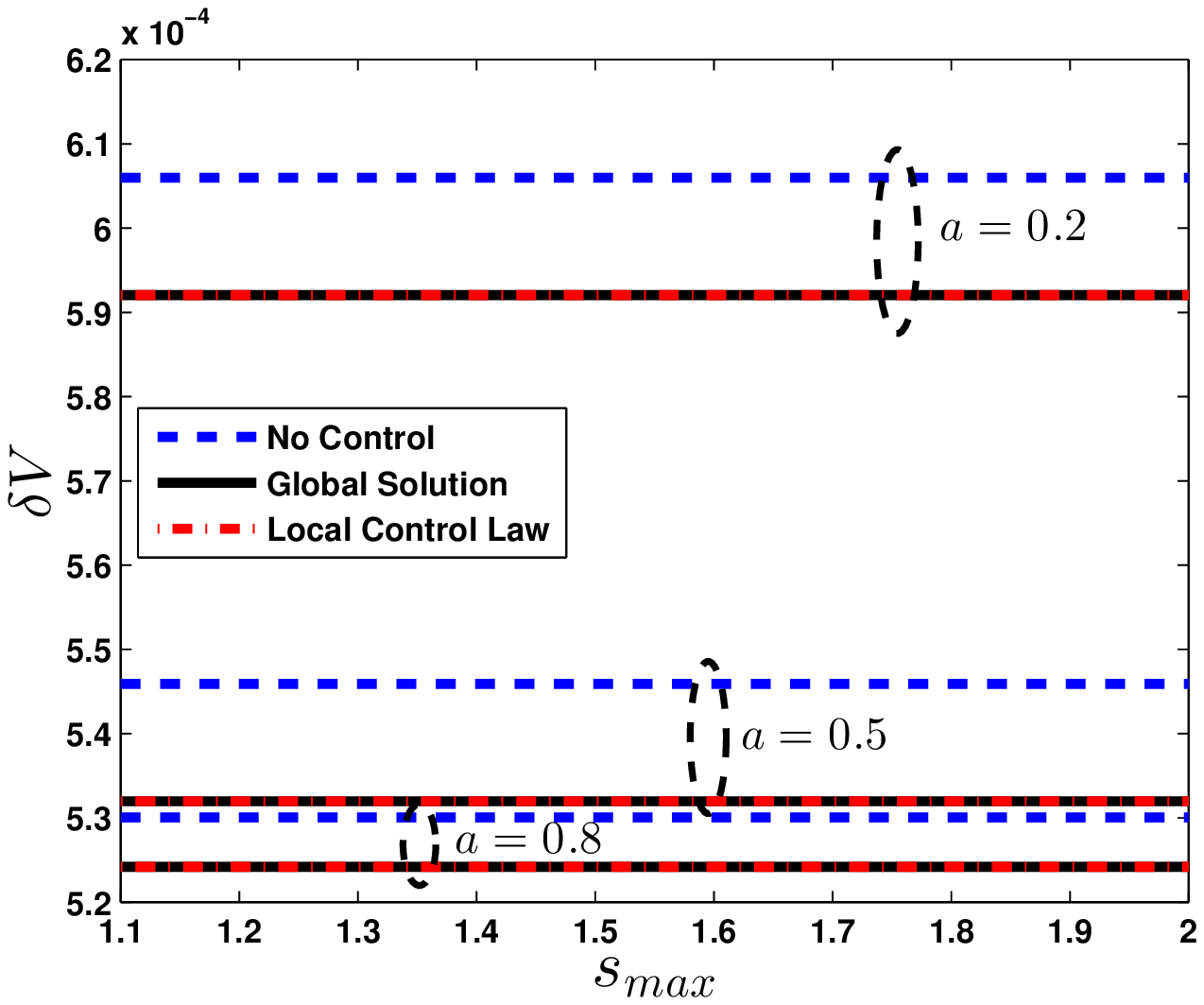}}
\caption{The maximum voltage variation $\delta V$ for (a) front battery placement and (b) rear battery placement as a function of $s_{max}$ at penetration levels $a=0.2$, $0.5$ and $0.8$. The figures provide a $\delta V$ comparison between front and rear battery placement for the global solution, local control law, and no control.}
\label{fig:dV_smax_compare}
\end{center}
\end{figure*}

Figure \ref{fig:dV_smax_compare} provides a comparison of voltage variation $\delta{V}$ between global optimal solution of the loss minimization problem, optimal local solution with local control law, and no control for front and rear battery placement. As expected, having no control produces the worst performance in all cases. Higher PV penetration also provides smaller voltage variations for both front and rear battery placements. One important observation can be made from Figure \ref{fig:dV_smax_compare} is that rear battery placement provides smaller voltage variations than front battery placement for all $s_{max}$ and all three levels of PV penetrations. In the case of rear battery placement and when $a=0.8$ for front battery placement, both global solution and local solution have the same voltage variation $\delta{V}$ for all  $s_{max}$. 



\begin{figure*}
\begin{center}
\subfigure[]{\label{fig:saving_smax_Front}
\includegraphics[height=2.45 in, width=3.3 in,clip]{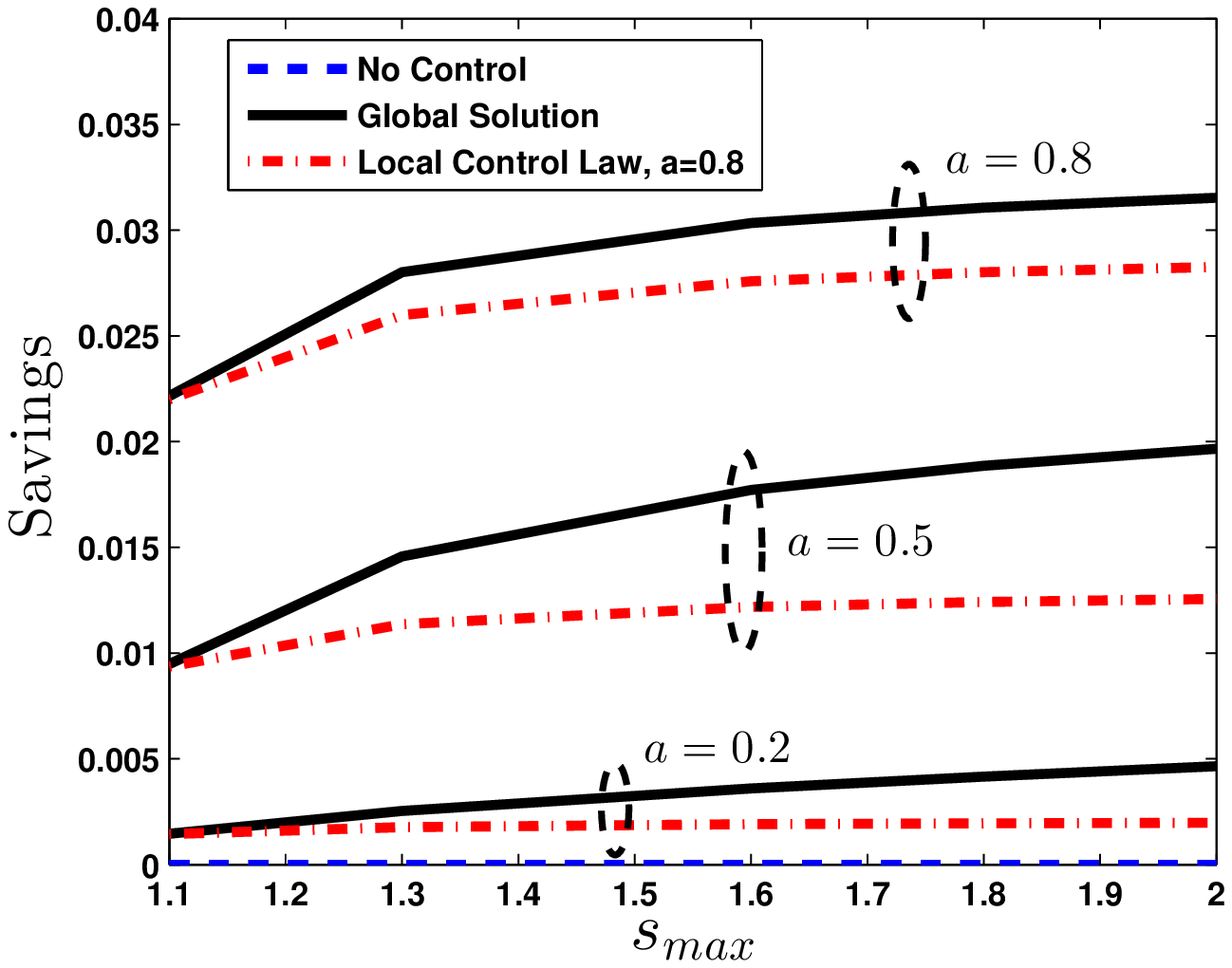}}
\subfigure[]{\label{fig:saving_smax_Rear}
\includegraphics[height=2.45 in, width=3.3 in,clip]{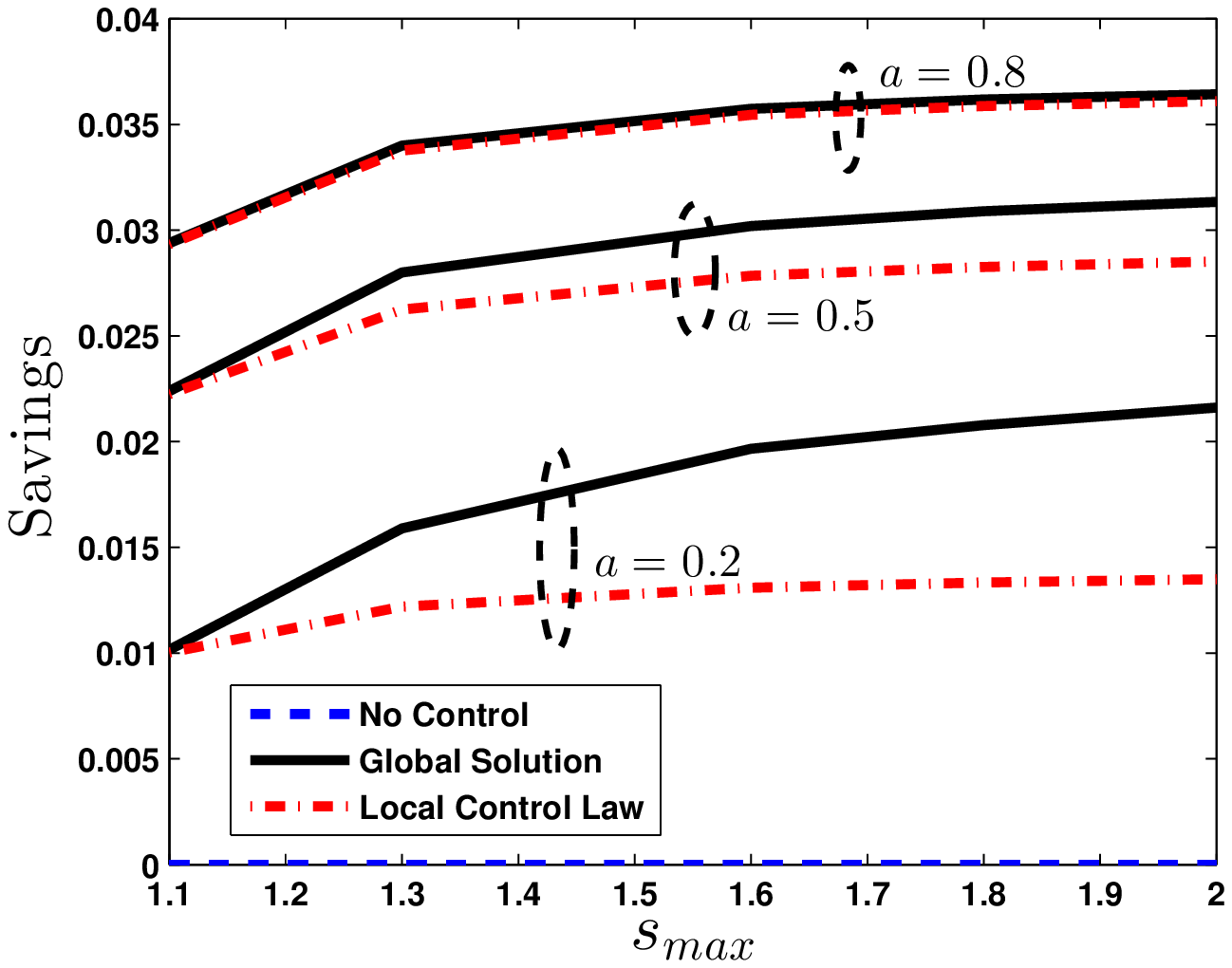}}
\caption{The energy saved for (a) front battery placement and (b) rear battery placement as a function of $s_{max}$ at penetration levels $a=0.2,\, 0.5$ and $0.8$. The figures provide an energy saving comparison between front and rear battery placement for the global solution, local control law, and no control.}
\label{fig:saving_smax_compare}
\end{center}
\end{figure*}

Figure \ref{fig:saving_smax_compare} provides a comparison of energy saving between global optimal solution of the loss minimization problem, optimal local solution with local control law, and no control for front and rear battery placement. As expected, having no control produces the worst performance in all cases. Higher PV penetration also offers higher energy saving for both front and rear battery placements. Similar important observation can be made from Figure \ref{fig:saving_smax_compare} is that rear battery placement provides higher energy savings than front battery placement for all $s_{max}$ and all three levels of PV penetrations. This is expected since rear battery placement provides smaller voltage variation than front battery placement as shown in Figure \ref{fig:dV_smax_compare}. Global solution works better than local control law in terms of energy saving. When the PV penetration level gets as high as $80\%$, the local control law provides equal energy saving as global solution for rear battery placement.

The maximum voltage variation $\delta V$ in the rear battery placement is smaller than in the front battery placement due to the fact that the front nodes are closer to the substation where the voltage is kept at a constant without requiring any voltage regulation. That's why $\delta{V}$ stays small at the front nodes as $V_j(t)$ isn't used as one of the optimizing variables. Moving toward the rear nodes, $V_j(t)$ is then used as one of the optimizing variables, which also causes $\delta{V}$ to become larger. As power demand increases and solar generation decreases at the peak time (i.e., $t = 50,...,65$) as shown in Figure \ref{fig:powerProfile}, the distribution circuit struggles to meet the high power demand with low PV generation and this behavior results in large voltage variations at the peak demand time in the rear nodes. In the case of the front battery placement during the peak demand time, we don't have any PV generation with battery that can provide control variables for regulating the voltage at the rear nodes; whereas in the case of rear battery placement, we have PV cells with battery at the rear nodes that can perform two tasks: 1) reduce the power demand needed at the rear nodes due to solar generation and energy storage in battery, 2) regulate the voltage by utilizing two control variables such as the charge/discharge rate $\beta_j(t)$ of the battery and the reactive power $q_j^g(t)$ of the inverters. Due to this reason, putting more PV cells with battery at the rear nodes will provide smaller voltage variations than at the front nodes. Hence, higher energy saving is also obtained with rear battery placement as shown in Figure \ref{fig:saving_smax_compare} for all PV penetration levels and $s_{max}$. For penetration level $a=0.2$ and $s_{max}=1.1$, note that the peak $\delta{V}$ in the rear battery placement is about $12\%$ smaller than the peak $\delta{V}$ in the front battery placement and the average $\delta{V}$ in the rear battery placement is about $5\%$ smaller than the  average $\delta{V}$ in the front battery placement, which results in an energy saving of $9\%$. As for medium PV penetration level of $a=0.5$ and $s_{max}=1.1$, the peak $\delta{V}$ in the rear battery placement is about $13\%$ smaller than the peak $\delta{V}$ in the front battery placement and the average $\delta{V}$  in the rear battery placement is about  $9\%$ smaller than the  average $\delta{V}$ in the front battery placement, which results in an energy saving of $15\%$.


\section{Conclusion and Future Research}
\label{sec:Conclusion}

This work is a continuation of our previous work to study the effects of energy storage (battery) placement on VAR control using dynamic control algorithms in a single line radial distribution system with $20\%$, $50\%$, and $80\%$ penetration of PV cells. Although the circuit model is unidirectional, we don't place any constraints on the direction of the power and allow it to be both positive or negative. Therefore, this model can be used in the case of bidirectional circuit. The dynamic control algorithms in this paper are designed to minimize the total power loss over one day. Although the number of nodes is limited to 30, it does show that the two dynamic control algorithms work much better than the no control case for both the front and rear battery placement. Battery placement at the rear of the distribution also offers smaller voltage variations and higher energy savings than at the front battery placement for all levels of PV penetration and sizes of inverters. The simulation results from this paper also suggests that if we were to install a number of systems of PV cells with battery, the good place to install these systems is at the rear of the distribution for low, medium, and high PV penetration levels as shown in both Figures \ref{fig:dV_smax_compare} and \ref{fig:saving_smax_compare}. This may provide an important message to utility companies to offer an "incentive" program to the users at the rear (not at the front) in the distribution circuits to install battery (more costly) in addition to install PV cells. This will improve the quality of service to all users in the distribution circuits. Our ongoing effort is to model the charging/discharging rate of the energy storage (battery) using physical constraints such as battery temperature rise and generated gas. Moreover, further analysis is needed to understand how the approach in this paper can be adapted to more general distribution system models and the effects of adding large numbers of PV-cells to a single phase of a real system. Additionally, the study of an optimal battery placement for a more general distribution system with a specific PV penetration level is needed.




\bibliographystyle{IEEEtr}  
\bibliography{OPF_BatteryPlacement}

\begin{IEEEbiography}[{\includegraphics[width=1in,height=1.25in,clip,keepaspectratio]{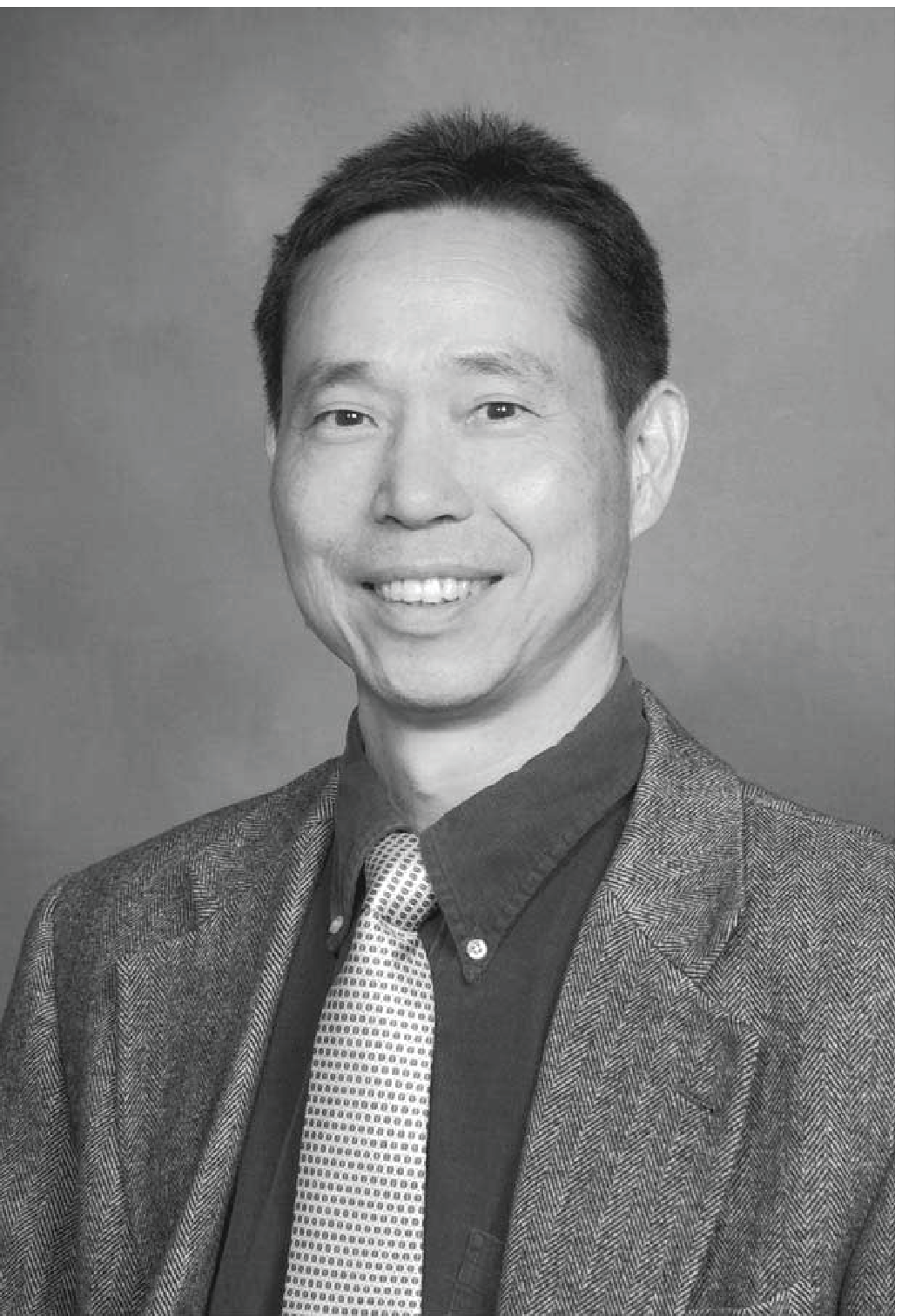}}]{Hen-Geul Yeh}
received the B.S. degree in engineering science from National Chen Kung University, Taiwan, in 1978, and the M.S. degree in mechanical engineering and the Ph.D. degree in electrical engineering from the University of California, Irvine, in 1979 and 1982, respectively.

He has been with the Electrical Engineering Department at California State University, Long Beach (CSULB), since 1983, as a Professor since 1986. His research activities have included smart grids, optimization and adaptive controls, and electrical event detection, real-time signal processing, wireless communication algorithms development and implementation. His areas of expertise are real-time DSP, Wi-Fi and Wi-MAX, adaptive systems, and mobile communication. He has been granted four U.S. patents.

\end{IEEEbiography}

\begin{IEEEbiography}[{\includegraphics[width=1in,height=1.25in,clip,keepaspectratio]{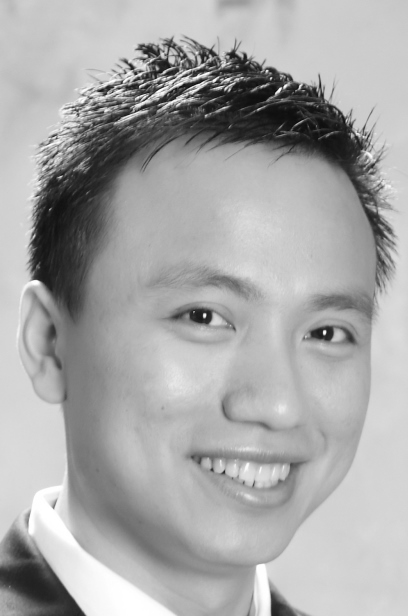}}]{Son H. Doan}
received the B.S.E.E. from the University of California, Los Angeles, in 2004, and the M.S.E.E. from the  University of Southern California in 2007. He is currently pursuing his Ph.D. in engineering mathematics.

\end{IEEEbiography}

\end{document}